\documentclass[aps,prd,twocolumn,amsmath]{revtex4-1}
\usepackage{color,epsfig,graphicx,amsmath,amssymb,float,multirow,xspace,times} 
\definecolor{dblue}{rgb}{0.00,0.00,0.75}
\usepackage[colorlinks,urlcolor=dblue,linkcolor=dblue,citecolor=dblue]{hyperref} 
\def \babar {{\it BABAR} }
\begin{document} 

\title{Subprocesses $\rho(770,1450)\to K\bar{K}$ for the three-body hadronic $D$ meson decays} 
                    
\author{Wen-Fei Wang$^{1,2}$\vspace{0.1cm}}\email{wfwang@sxu.edu.cn}

\affiliation{$^1$Institute of Theoretical Physics, Shanxi University, Taiyuan, Shanxi 030006, China \\
                 $^2$State Key Laboratory of Quantum Optics and Quantum Optics Devices, Shanxi University, 
                        Taiyuan, Shanxi 030006, China  \vspace{0.2cm} }

\date{\today}

\begin{abstract}
We construct the theoretical framework for quasi-two-body $D$ meson decays with the help of pion and kaon electromagnetic 
form factors, and with which we study the contributions of the subprocesses $\rho(770,1450)\to K\bar{K}$ for the three-body 
$D$ decays within the flavour \emph{SU}(3) symmetry. Because of the limitations imposed by phase space and strong coupling, 
the contributions for kaon pair from the virtual bound state $\rho(770)$ are channel-dependent and generally small for the 
concerned three-body $D$ decays, but some quasi-two-body processes could still be observed in the Dalitz plot analyses 
for related decays, such as $D^0 \to K^-\rho(770)^+ \to K^- K^+K_S^0$ and $D^+ \to K_S^0\rho(770)^+ \to K_S^0 K^+K_S^0$, 
they are predicted to have the branching fractions $\mathcal{B}=(0.82\pm0.04)\times 10^{-4}$ and 
$\mathcal{B}=0.47^{+0.05}_{-0.03}\times 10^{-4}$, which are $(1.86\pm0.16)\%$ and  $(1.84^{+0.21}_{-0.16})\%$, respectively, 
of the total branching fractions for the corresponding three-body $D$ decays. We find in this work that the normal subprocesses 
like $\rho(1450)^+\to \pi^+\pi^0$ or $\rho(1450)^+\to K^+\bar{K}^0$, which are bound by the masses of decaying initial states, 
will provide virtual contributions in some special decays. 
\end{abstract}

\maketitle

\section{Introduction}
Three-body hadronic $D$ meson decays provide us a rich field to test the Standard Model and beyond. These decays are 
proceeded predominantly through the quasi-two-body processes~\cite{zpc34-103,prd75-052003,prd79-032003}. 
Due to their small nonresonant components and abundant intermediate states, three-body $D$ decays as well as their 
subprocesses were widely employed to study the properties and substructures of various resonant states~\cite{plb502-79,
prl86-765,prl86-770,plb579-59,plb632-471,plb651-129,plb742-363,prd94-096002,prd96-036013,epjc80-1041,epjc80-895,
plb803-135279,prd103-034020,2102-05349}, to analyses hadron-hadron interactions~\cite{prd92-094005,prd92-054010,
JHEP1408-135,prd84-094001,prd80-054007,prd71-054030,prd68-094012}, and to extract information on the $\pi\pi$, 
$K\pi$, and $KK$  $S$-wave amplitude in the low energy region~\cite{2011-08041,prd83-052001,prl89-121801,plb681-14,
prd78-052001,plb653-1,plb648-156,prd76-012001,prd76-011102,prd73-032004,prd72-031102}.  
In the experimental analyses for relevant decay amplitudes~\cite{2103-15098,2006-02800,prd102-012002,prl123-112001,
prd93-052018,prd89-052001,prd84-092005,prd83-052001}, Dalitz plot technique~\cite{dalitz} was widely adopted in recent years.
The corresponding expressions of the decay amplitudes are usually composed of coherent sum of the 
resonant and nonresonant contributions within the isobar formalism~\cite{pr135-B551,pr166-1731,prd11-3165}.  
For the precise and accurate Dalitz plot analyses, all reliable and necessary strong dynamical components should present in the 
expressions of the decay amplitudes. 

The contributions for kaon pair in the final states of three-body $D$ decays from the $\rho$ family resonances, such as 
$\rho(1450)^{\pm}$ and $\rho(1700)^{\pm}$ for $K^0_S K^{\pm}$ in the decays $D^+\to K^0_SK^+\pi^0$ and 
$D^0\to K^0_S K^{\pm}\pi^{\mp}$, have been noticed by BESIII~\cite{2104-09131}, LHCb~\cite{prd93-052018} and 
CLEO~\cite{prd85-092016} collaborations. In addition, the subprocess $\rho(1450)^0\to K^+K^-$ was found 
contributed a surprising large fit fraction for the three-body decays $B^\pm\to \pi^\pm K^+K^-$ by LHCb collaboration in 
Ref.~\cite{prl123-231802}. The results of Ref.~\cite{prd103-114028} indicates a sizeable branching fraction, around $22\%$, 
of $D^0\to K^0_S K^+K^-$ comes from the subprocesses $\rho(770,1450,1700)^+\to K^0_S K^+$. 
For the quasi-two-body $B$ meson decays, resonance contributions for the kaon pair originating from the intermediate states 
$\rho(770,1450,1700)$ have been specifically studied in Refs.~\cite{prd101-111901,prd103-056021,prd103-016002}. 
In this paper, we shall extend our previous works~\cite{prd101-111901,prd103-056021,prd103-016002} to the quasi-two-body 
$D$ meson decays and concentrate on the contributions of $\rho(770,1450)\to K\bar{K}$ for relevant processes.

\begin{figure}[tbp]  
\centerline{\epsfxsize=8cm \epsffile{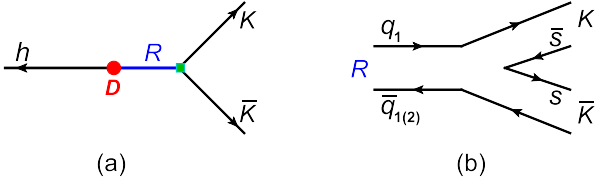}}
\caption{Schematic view of the cascade decays $D\to hR \to h K\bar{K}$, where $R$ stands for the intermediate states 
              $\rho(770,1450)$ decaying into $K\bar{K}$ in this work, and $q_{1,2}$ is the light quark $u$ or $d$. }
\label{fig-1}
\end{figure}

The schematic diagram for the cascade decays $D\to hR \to h K\bar{K}$ is shown in Fig.~\ref{fig-1}.  In the rest frame of the 
initial state, $D$ meson decays into the intermediate state $R$ and the bachelor $h$, and then $R$ decays into its daughters 
the kaon pair. When $q_{1,2}$ is the light quark $u$ or $d$ as shown in the Feynman diagram Fig.~\ref{fig-1}-(b), the intermediate 
states could be the resonances $\rho(770,1450)$.
The natural decay modes for $\rho(770)\to K\bar{K}$ are blocked because the pole mass for resonant state $\rho(770)$ is below 
the threshold of kaon pair. However, the virtual contribution~\cite{plb25-294,Dalitz62,prd94-072001,plb791-342} from the 
Breit-Wigner (BW) formula~\cite{BW-model} tail of $\rho(770)$ was found to be indispensable for the production of kaon pair in 
the  processes of $\pi^-p\to K^-K^+n$ and $\pi^+n\to K^-K^+p$~\cite{prd15-3196,prd22-2595}, $\bar p p \to K^+K^-
\pi^0$~\cite{plb468-178,epjc80-453}, $e^+e^- \to K^+K^-$~\cite{pl99b-257,pl107b-297,plb669-217,prd76-072012,zpc39-13,
prd88-032013,prd94-112006,plb779-64,prd99-032001} and $e^+ e^- \to K^0_{S}K^0_{L}$~\cite{pl99b-261,
prd63-072002,plb551-27,plb760-314,prd89-092002,jetp103-720}. Besides, $\rho(770,1450)^\pm$ are important intermediate 
states for $K^\pm K^0_S$ for the final states of hadronic $\tau$ decays~\cite{prd98-032010,prd89-072009,prd53-6037,epjc79-436}. 
With the kaon electromagnetic form factors studied in 
Refs.~\cite{epjc79-436,zpc39-13,prd76-072012,prd88-032013,prd67-034012,epjc39-41,epjc49-697,prd81-094014,jetp129-386} 
including the $\rho, \omega$ and $\phi$ families resonant states, we predicted the branching fractions for the 
charmless decays $B\to h\rho(1450) \to h K\bar{K}$ (with $h=\pi, K$) to be about a tenth of their corresponding quasi-two-body 
decays with $\rho(1450)$ decaying into pion pair~\cite{prd101-111901,prd103-056021}. What's more, we found that 
the branching fraction of virtual contribution for $K\bar K$ from the BW tail of $\rho(770)$ is larger than the corresponding 
contribution from $\rho(1450)$ in a charmless quasi-two-body $B$ decay~\cite{prd103-056021}.

Unlike hadronic $B$ meson decays, for which the heavy quark expansion tools and factorization approaches have been successfully 
used for decades, the two-body and three-body hadronic $D$ meson decays are challenging to be reliably described based on 
quantum chromodynamics (QCD) on theoretical side because of the $c$-quark mass. In this context, model independent ways  %
such as the factorization-assisted topological-amplitude approach~\cite{prd86-036012,prd89-054006,adv-7627308} and          
the topological-diagram approach~\cite{prd85-054014,prd85-034036,prd86-014014,prd93-114010,prd100-093002,2104-13548} 
have been adopted in references for various hadronic $D$ decays. In addition, starting from the weak effective 
Hamiltonian~\cite{rmp68-1125},  the experimental data for the decays $D^0\to K^0_S K^+K^-$ from 
\babar~\cite{prl105-081803} and $D^0\to K^0_S \pi^+\pi^-$ from \babar~\cite{prl105-081803} and Belle~\cite{prl99-131803} 
were analyzed in~\cite{prd103-114028} and~\cite{prd89-094018}, respectively, within the quasi-two-body factorization framework.
While in~\cite{epjc81-268}, with the chiral unitary approach be used to take into account the final state interaction, 
a model was developed to study the three-body $D^0\to K^-\pi^+\eta$ decay. In~\cite{prd98-056021}, the multimeson 
model based on chiral effective Lagrangians was proposed  to describe the $D^+\to K^- K^+K^+$ decay.
The decay $D^+\to K^-\pi^+\pi^+$ was studied in~\cite{JHEP1510-142} utilizing dispersion theory. 
The $\Delta U=0$ rule in three-body charm decays was studied in Ref.~\cite{jhep2105-179}. 
Flavor $SU(3)$ sum rules for $D \to PP$ and $D \to PV$ decay amplitudes were presented in~\cite{JHEP1304-067}, 
where $P$ is a light pseudoscalar and $V$ is a light vector. And $SU(3)$ flavor symmetry relationship was used in 
Ref.~\cite{prd82-114032} to analyze the ratios of amplitudes and phases for $D^0\to PV$ in the decays 
$D^0\to \pi^+\pi^-\pi^0$, $D^0\to \pi^0 K^+K^-$ and $D^0\to K^-\pi^+\pi^0$.

This paper is organized as follows: In Sec.~\ref{sec-2}, we construct the theoretical framework for the quasi-two-body $D$ meson decays with the help of the pion and kaon electromagnetic form factors, we derive the relation between the branching fractions for 
the two-body and quasi-two-body $D$ meson decays.  In Sec.~\ref{sec-3}, we present our numerical results for the contributions 
of the subprocesses $\rho(770,1450)\to K\bar{K}$ for concerned $D$ meson decays, along with some necessary discussions.
A brief summary of this work is given in Sec.~\ref{sec-sum}.

\section{Framework}\label{sec:2}
\label{sec-2}

For the cascade decays $D\to hR\to h h_1h_2$, with $h_{1(2)}$ is a light pseudoscalar pion or kaon, the related effective 
weak Hamiltonian is written as~\cite{rmp68-1125}
\begin{eqnarray}
\mathcal{H}_{\rm eff} = \frac{G_F}{\sqrt2}&\bigg[& \sum_{q=d,s} \lambda_q (C_1 O_1+ C_2 O_2)                        \nonumber \\
                                                                   & - & \lambda_b \sum^{6}_{i=3}C_i O_i  -  \lambda_b C_{8g} O_{8g}\bigg],
\label{eq-hamilton}
\end{eqnarray}
where $G_F=1.1663787(6)\times10^{-5}$ GeV$^{-2}$~\cite{PDG-2020} is the Fermi coupling constant, $\lambda_q=V^*_{cq}V_{uq}$ 
and $\lambda_b=V^*_{cb}V_{ub}$ stand for the product of the Cabbibo-Kobayashi-Maskawa (CKM) matrix elements, $C$'s are the 
Wilson coefficients at scale $\mu$, the $O_1$ and $O_2$ are current-current operators, $O_3$-$O_4$ are QCD penguin operators, 
and $O_{8g}$ is chromomagnetic dipole operator. With Eq.~(\ref{eq-hamilton}) the effective weak Hamiltonian, the decay amplitudes for 
two-body processes $D\to PV$ were described with typical topological diagram amplitudes $T_{P,V}, C_{P,V}, E_{P,V}$ and $A_{P,V}$ 
as well as additional penguin amplitudes in the factorization-assisted topological-amplitude approach~\cite{prd89-054006} and the 
topological-diagram approach~\cite{prd85-034036,prd93-114010,prd100-093002,2104-13548}. For the detailed discussions 
on these the topological diagram amplitudes one is referred to the Refs.~\cite{prd89-054006,prd85-034036,prd93-114010,
prd100-093002,2104-13548}. Take $D^0\to K^- \rho(770,1450)^+ \to K^- \pi^+\pi^0 $ and 
$D^0\to K^-\rho(770,1450)^+ \to K^- K^+\bar{K}^0 $ as the examples, now we construct the decay amplitudes for the decays 
$D\to hR\to h h_1h_2$.

If the subprocesses $\rho(770,1450) \to \pi\pi$ were shrunk to the mesons $\rho(770,1450)$, the quasi-two-body processes
$D^0\to K^- \rho(770,1450)^+ \to K^-\pi^+\pi^0 $ will become the two-body channels $D^0\to K^- \rho(770,1450)^+ $. The decay 
amplitudes of $D^0\to K^- \rho(770,1450)^+ $ are dominated by the color-favored tree amplitude $T_P$ with $D^0\to K^-$ transition, 
which is formulated as~\cite{prd85-034036,prd89-054006,prd93-114010,prd100-093002}
\begin{eqnarray} 
  T_P &=& \frac{G_F}{\sqrt2} V^*_{cs}V_{ud} \left[ \frac{C_1}{3}+C_2 \right] f_\rho m_\rho  \nonumber\\
                   &\times&  F_1^{D\to K}(m^2_\rho) 2 \left[\epsilon_\rho\!\cdot\! p_D\right], 
  \label{def-Tp}
\end{eqnarray}  
where the subscript $\rho$ stands for $\rho(770)$ or $\rho(1450)$, $F_1^{D\to K}$ is the form factor for $D^0\to K^-$ transition
and is parametrized as~\cite{epjd4-2}
\begin{eqnarray}
     F_1^{D\to K}(Q^2) = \frac{0.78}{(1 - Q^2 / 2.11^2)(1 - 0.24 Q^2 / 2.11^2)}\,, 
\end{eqnarray}
$\epsilon_\rho$ is the polarization vector, and $p_D$ is the momentum for $D^0$. Beyond the amplitude $T_P$ there is a
W-exchange nonfactorizable contribution from amplitude $E_V$, which is extracted from experimental data in topological-diagram 
approach~\cite{prd85-034036,prd93-114010,prd100-093002} and is parametrized as~\cite{prd89-054006}
\begin{eqnarray} 
  E_V &=& \frac{G_F}{\sqrt2} V^*_{cs}V_{ud} C_2 \chi^E_q e^{i \phi^E_q} f_D m_D  \left[ f_\rho/f_{\rho(770)} \right]  \nonumber\\
                   &\times&   \left[ f_K/f_\pi \right]  \left[ \epsilon_\rho\!\cdot\! p_D \right],
  \label{def-Ev}
\end{eqnarray}
in the factorization-assisted topological-amplitude approach, with the decay constants $f_D$, $f_K$ and $f_\pi$ for the $D^0$, 
kaon and pion, respectively, and the $\chi^E_q=0.25$ and $\phi^E_q=1.73$~\cite{prd89-054006} are the parameters characterize 
the strength and strong phase of the corresponding amplitude. Thus, one has the decay amplitudes $\mathcal{M}=T_P+E_V$ 
for the decays $D^0\to K^-\rho(770,1450)^+$.  Utilizing the partial decay width~\cite{PDG-2020} 
\begin{eqnarray} 
   \Gamma(D\to \rho K)=\frac{\big| \overrightarrow{p}\big|^3}{8\pi m^2_\rho}\big|\widetilde{\mathcal{M}}\big|^2,
   \label{def-dew-2B}       
\end{eqnarray}
and the mean life of $D^0$ meson, one will achieve the branching fractions for the two-body decays $D^0\to K^- \rho(770,1450)^+$ 
with the relation $\mathcal{M}=\epsilon_\rho\! \cdot\! p_D\widetilde{\mathcal{M}}$, where the magnitude of the momentum for 
$\rho$ or $K$ in the rest frame of $D$ meson is
\begin{eqnarray}
    &&\left| \overrightarrow{p} \right|=   \nonumber\\
          &&\;\;\; \frac{1}{2 m_D} \sqrt{\left[m^2_{D}-(m_\rho+m_{K})^2\right]\left[m^2_{D}-(m_\rho-m_{K})^2\right]}, \;\;
    \label{def-2B-p}
\end{eqnarray}  
and the $m_{i}$'s are the masses for relevant particles above with $i=\{D, \rho, K\}$.

The subprocesses $\rho(770,1450)^+ \to \pi^+\pi^0$ of the quasi-two-body decays $D^0\to K^-\rho(770,1450)^+\to K^-\pi^+\pi^0$ 
are associated with the pion electromagnetic form factor $F_\pi(s)$, with $s$ the squared invariant mass for pion pair. The form 
factor for pion has been measured with high precision in Refs.~\cite{plb753-629,plb720-336,prd86-032013,prd78-072006,
plb648-28,pr421-191,prd61-112002} by different collaborations. For the $\rho(770)$ and $\rho(1450)$ components, one 
has~\cite{epjc39-41,prd81-094014}
\begin{eqnarray} 
   F_\pi^{R}(s)=c^\pi_{R} {\rm BW}_{R}(s),
   \label{def-Fpi}
\end{eqnarray}
where $R$ represents the resonance $\rho(770)$ or $\rho(1450)$, the coefficient 
$c^\pi_{R}=f_R g_{R\pi\pi}/(\sqrt2 m_R)$~\cite{epjc39-41} is determined by the decay constant $f_R$, coupling constant 
$g_{R\pi\pi}$ and the mass $m_R$. The BW formula for $F_\pi$ has the form~\cite{zpc48-445}
\begin{eqnarray}
   {\rm BW}_R= \frac{m_{R}^2}{m_R^2-s-i m_R \Gamma_{R}(s)}\,,    
   \label{eq-BW}
\end{eqnarray}
and the $s$-dependent width is     
\begin{eqnarray}\label{def-width}
 \Gamma_{R}(s)
             =\Gamma_R\frac{m_R}{\sqrt s} \frac{ \left| \overrightarrow{q} \right|^3}{ \left| \overrightarrow{q_0}\right|^3} 
                X^2(\left| \overrightarrow{q} \right| r^R_{\rm BW}).
  \label{eq-sdep-Gamma}
\end{eqnarray}
The Blatt-Weisskopf barrier factor for $\rho$ family resonances is given by~\cite{BW-X}  
\begin{eqnarray}
     X(z)=\sqrt{\frac{1+z^2_0}{1+z^2}}\,,
\end{eqnarray}
with the barrier radius $r^R_{\rm BW}=1.5$ GeV$^{-1}$~\cite{prd63-092001}.
The magnitude of the momentum $\left| \overrightarrow{q} \right|=\frac{1}{2}\sqrt{s-4m_\pi^2}$ in Eq.~(\ref{eq-sdep-Gamma}), and 
$\left| \overrightarrow{q_0}\right|$ is $\left| \overrightarrow{q} \right|$ at $s=m^2_R$. 

By connecting subprocesses $\rho(770,1450)^+ \to \pi^+\pi^0$ and the two-body channels $D^0\to  K^-\rho(770,1450)^+$ together, 
we get the quasi-two-body decay processes. The amplitudes for $D^0\to K^- \rho(770,1450)^+\to K^- \pi^+\pi^0$ are written as 
\begin{eqnarray} 
                       && \langle (\pi^+\pi^0)_\rho K^- |\mathcal{H}_{\rm eff}| D^0 \rangle                                                    \nonumber \\
                       && \quad=\langle \pi^+\pi^0| \rho(770,1450)^+ \rangle \frac{1}{m_R^2-s-i m_R \Gamma_{R}(s)}       \nonumber \\ 
                       && \quad\times\langle \rho(770,1450)^+ K^- |\mathcal{H}_{\rm eff}| D^0 \rangle                                \nonumber \\
                       && \quad= {g_{\rho\pi\pi} \epsilon_\rho \cdot  (p_{\pi^+}-p_{\pi^0})}
                                    \frac{1}{m_\rho^2-s-i m_\rho \Gamma_{\rho}(s)}                                                                      \nonumber \\
                       && \quad\times  \left[ T_P+E_V \right],
    \label{def-amp-Q2B}
\end{eqnarray}  
where the $p_{\pi^+}$ and $p_{\pi^0}$ are the four momenta for $\pi^+$ and $\pi^0$, respectively. Then in the rest frame of the 
intermediate states, we have the expression of the differential branching fraction ($\mathcal{B}$)~\cite{PDG-2020} 
\begin{eqnarray} 
   \frac{d{\mathcal B}}{\sqrt{s}\, d\sqrt{s}}=\tau_D\frac{\left| \overrightarrow{q} \right|^3 \left| \overrightarrow{p_h}\right|^3}
                                                                 {12\pi^3m^3_D} \big|\mathcal{A} \big|^2\;,
    \label{eqn-diff-bra}  
\end{eqnarray}
by taking into account the Eq.~(\ref{def-Fpi}), with $\tau_D$ the mean lifetime for $D$ meson. 
The magnitude of the momentum ${\small|\overrightarrow{p_h}|}$ for the state $h$  is written as
\begin{eqnarray}    
   &&\left| \overrightarrow{p_h}\right|  =                         \nonumber \\
     && \quad\frac{1}{2\sqrt s} \sqrt{\left[m^2_{D}-(\sqrt{s}+m_{h})^2\right]\left[m^2_{D}-(\sqrt{s}-m_{h})^2\right]}, \;\;\; 
                \label{def-qh}
\end{eqnarray}  
with $m_h$ the mass for the bachelor final state, which is $K^-$ for $D^0\to K^- \rho(770,1450)^+\to K^-\pi^+\pi^0$ decays. The 
decay amplitude $\mathcal{A}$ in Eq.~(\ref{eqn-diff-bra}) is related to that of the corresponding two-body process with the relation
\begin{eqnarray} 
      \mathcal{A}\approx \widetilde{\mathcal{M}}/m_\rho,
\end{eqnarray}
and with the replacement $f_\rho\to F^\rho_\pi(s)$ for the resonant states. 

The subprocesses $\rho(770,1450)^+ \to K^+\bar{K}^0$ in the decays $D^0\to K^- \rho(770,1450)^+\to K^- K^+\bar{K}^0$ are 
related to the kaon electromagnetic form factors. These form factors have been extensively studied in Refs.~\cite{epjc79-436,
prd67-034012,epjc39-41,epjc49-697,prd81-094014} on the theoretical side. The experimental information on kaon form factors 
comes from the measurements of the reactions $e^+e^- \to K^+K^-$~\cite{zpc39-13,prd76-072012,prd99-032001} and 
$e^+e^- \to K^+K^-(\gamma)$~\cite{prd88-032013}.  Since $K\bar K$ is not an eigenstate of isospin, both isospin $0$ and 
$1$ resonant states such $\phi,~\omega$ and $\rho$ families need to be taken into the components of the form factors for 
kaon~\cite{prd88-032013}.  When concern only the contributions for ${K^+K^-}$ and ${K^0\bar K^0}$ from the resonances 
$\iota=\rho(770,1450,1700)$ and $\varsigma=\omega(782,1420,1650)$, one has~\cite{prd72-094003} 
\begin{eqnarray} 
   &&F_{K^+K^-}^{u}(s)=F_{K^0\bar K^0}^{d}(s)                                                                    \nonumber\\
     && \quad =+ \frac12\sum_{\iota} c^K_\iota {\rm BW}_\iota(s) 
                    + \frac12\sum_{\varsigma} c^K_\varsigma {\rm BW}_\varsigma(s),   \label{def-F-u}  \\
   &&F_{K^+K^-}^{d}(s)=F_{K^0\bar K^0}^{u}(s)                                                                     \nonumber\\
      &&\quad =- \frac12\sum_{\iota} c^K_\iota {\rm BW}_\iota(s) 
                    + \frac12\sum_{\varsigma} c^K_\varsigma {\rm BW}_\varsigma(s).   \label{def-F-d}
\end{eqnarray}
For the $K^+\bar K^0$ and $K^0K^-$ pairs which have no contribution from the neutral states $\omega(782,1420,1650)$, 
one has~\cite{prd67-034012,epjc39-41,epjc79-436}
\begin{eqnarray} 
  F_{K^+\bar K^0}(s)=F_{K^0K^-}(s)=\sum_{\iota} c^K_\iota {\rm BW}_\iota(s).   
  \label{def-F-ud}
\end{eqnarray}
The relevant coefficients $c^K_R$'s (with $R=\iota,\varsigma$) in kaon form factors, which are proportional to their corresponding 
coupling constants $g_{R K\bar K}$, have been detailed discussed in our previous work~\cite{prd103-056021}, we adopt the same 
values for them as in Ref.~\cite{prd103-056021}. With the relations $g_{\rho(770)K\bar K}=g_{\omega(782)K\bar K}=
\frac{1}{\sqrt2}g_{\phi(1020)K\bar K}$ and $g_{\rho(1450)K \bar K}\approx \frac12 g_{\rho(1450)\pi^+ \pi^-}$ within flavour $SU(3)$ 
symmetry~\cite{epjc39-41} and the Eq.~(\ref{def-amp-Q2B}) with the replacements $\pi^+\to K^+$ and $\pi^0\to \bar K^0$, 
one could easily obtain the decay amplitudes for the cascade decays $D^0\to \rho(770,1450)^+ K^-\to K^+\bar{K}^0 K^-$.

By neglecting the effect of the alteration of $m^2_\rho \to s$ for $F_1^{D\to K}$ in Eq.~(\ref{def-Tp}), one can relate partial decay 
width expression the Eq.~(\ref{def-dew-2B}) to Eq.~(\ref{eqn-diff-bra}) the differential branching fraction for quasi-two-body decays. 
From Eq.~(\ref{def-dew-2B}) we have 
\begin{eqnarray} 
    \big| \widetilde{\mathcal{M}}/m_\rho \big|^2 = \mathcal{B} (D\to \rho K)  
                  \frac{1}{\tau_D}\frac{8\pi}{\big| \overrightarrow{p}\big|^3}\,.
  \label{eq-DA2B}
\end{eqnarray}
The right hand side of this equation can be undoubtedly transplanted to the differential branching fraction of Eq.~(\ref{eqn-diff-bra}) 
along with the replacements $f_\rho\to F^\rho_{\pi,K}(s)$ for the quasi-two-body decays. Then we reach the branching fraction
\begin{eqnarray} 
    \mathcal{B}_{\rm Q2B}\approx \mathcal{B} (D\to \rho K)  \int\! d s \frac{\left| \overrightarrow{q} \right|^3}{3\pi^2m^3_D}
                       \left| \frac{\overrightarrow{p_h}}{\overrightarrow{p}} \right|^3  \left| \frac{F^\rho_{\pi,K}(s)}{f_\rho} \right|^2 \;\;\;
     \label{rela-2B-Q2B}    
\end{eqnarray}
for the quasi-two-body decays with the subprocesses $\rho(770)\to \pi\pi$ or $\rho(770)\to K\bar K$. This formula 
for the branching fractions of the quasi-two-body decays with the corresponding two-body results is different from the narrow 
width approximation relation discussed in Ref.~\cite{plb813-136058}, which is not appropriate for the processes with the virtual 
bound state~\cite{plb25-294,Dalitz62} decays like $\rho(770)\to K\bar K$. The integral part of Eq.~(\ref{rela-2B-Q2B}) is 
approximately equal the branching fraction for the normal subprocess such as $\rho(770)\to \pi\pi$ decay, but for the virtual 
processes like $\rho(770)\to K\bar K$, its integral result could be different for each decay channel.

\section{Results and Discussions}\label{sec:3}
\label{sec-3}

In the numerical calculation, we adopt the decay constant $f_{\rho}=0.216\pm0.003$ GeV~\cite{jhep1608-098} for $\rho(770)$, 
and the mean lives $\tau_{D^\pm}=1.040(7)$ ps, $\tau_{D^0}=0.4101(15)$ ps and $\tau_{D^\pm_s}=0.504(4)$ ps~\cite{PDG-2020} 
for $D_{(s)}$ mesons. For $\rho(1450)$, we employ $f_{\rho(1450)}=0.185^{+0.030}_{-0.035}$ GeV~\cite{plb763-29} resulting 
from the data~\cite{zpc62-455}. The masses for particles in relevant decay processes, the decay constants for pion and 
kaon, the full widths for resonances $\rho(770)$ and $\rho(1450)$ (in units of GeV), the Wolfenstein parameters for CKM matrix 
elements~\cite{PDG-2020}, and the decay constants $f_{D}$ and $f_{D_s}$ for $D_{(s)}$~\cite{epjc80-113} are presented 
in Table~\ref{tab1}.

\begin{table}[thb]  
\begin{center}
\caption{Masses for relevant states, decay constants for pion and kaon, full widths of $\rho(770)$ and $\rho(1450)$ 
              (in units of GeV) and the Wolfenstein parameters for CKM matrix elements from~\cite{PDG-2020},  and $f_{D}$ and 
              $f_{D_s}$ in~\cite{epjc80-113}.}
\label{tab1}
\begin{tabular}{c c c c }\hline\hline
  $m_{D^{\pm}}=1.870$     & $m_{D^{0}}=1.865$           & $m_{D^{\pm}_s}=1.968$       & $m_{\pi^{\pm}}=0.140$       \\
  $m_{\pi^{0}}=0.135$        & $m_{K^{\pm}}=0.494$       &  $m_{K^{0}}=0.498$              & $f_{D}=0.212$                       \\  
   $f_{D_s}=0.250$              &  $f_{\pi}=0.130$                 & $f_{K}=0.156$\;                     &                                               \\  
 \hline                
  \multicolumn{2}{l}{\;$m_{\rho(770)}=0.775$ }                      &  \multicolumn{2}{r}{$\Gamma_{\rho(770)}=0.149$\;}                      \\
  \multicolumn{2}{l}{\;$m_{\rho(1450)}=1.465\pm0.025$ }     &  \multicolumn{2}{r}{$\Gamma_{\rho(1450)}=0.400\pm0.060$\;}    \\
  \multicolumn{2}{l}{\;$\lambda=0.22650\pm 0.00048$}        &  \multicolumn{2}{r}{$A=0.790^{+0.017}_{-0.012}$\;}                     \\
  \multicolumn{2}{l}{\;$\bar{\rho} = 0.141^{+0.016}_{-0.017}$ }       &  \multicolumn{2}{r}{$\bar{\eta}= 0.357\pm0.011$\;}            \\
 \hline\hline   
\end{tabular}
\end{center}
\end{table}

Utilizing Eq.~(\ref{eqn-diff-bra}) the differential branching fraction, we have the branching fraction  
\begin{eqnarray} 
  & & \mathcal{B}(D^0 \to K^-\rho(770)^+ \to K^- \pi^+\pi^0)  \nonumber\\
  & & \hspace{1cm}    = (9.40\pm0.26\pm0.21) \%,           
   \label{res-K-R+}        
\end{eqnarray}
with the amplitudes $T_P$ and $E_V$ in Eq.~(\ref{def-Tp}) and Eq.~(\ref{def-Ev}), respectively. The coefficient $c^\pi_\rho=1.177$ 
for the pion form factor with 
\begin{eqnarray} 
    g_{\rho(770)\pi\pi}=\sqrt{\frac{6\pi m^2_{\rho(770)}\Gamma_{\rho(770)} }{\left| \overrightarrow{q_0}\right|^3}}
\end{eqnarray}
has been adopted. The two errors of the result (\ref{res-K-R+}) are induced by the uncertainties of the decay constant $f_{\rho(770)}$ 
and the CKM matrix elements,  respectively. The quasi-two-body branching fraction Eq.~(\ref{res-K-R+}) is consistent with the result 
$\mathcal{B}=9.6\%$ in Ref.~\cite{prd89-054006} within factorization-assisted topological-amplitude approach for the two-body 
decay $D^0 \to K^-\rho(770)^+$ in view of $\mathcal{B}(\rho(770) \to \pi\pi)\sim100\%$~\cite{PDG-2020}, but slightly less than 
the data $\mathcal{B}=(11.3\pm0.7)\%$ in {\it Review of Particle Physics}~\cite{PDG-2020} averaged from~\cite{prd63-092001,
plb331-217}. While with the form factor $F_\pi(s)$ measured in~\cite{prd86-032013} by \babar collaboration with Gounaris-Sakurai 
(GS) model~\cite{GS-model} for the $\rho$ family resonances, the branching fraction will be enhanced to be 
$\mathcal{B}=(10.19\pm0.28\pm0.23)\%$ for the quasi-two-body decay $D^0 \to K^-\rho(770)^+ \to K^- \pi^+\pi^0$.

Now we switch to the subprocess $\rho(770)^+\to K^+\bar{K}^0$ for $D^0 \to K^-\rho(770)^+$ decay.  Because of 
the pole mass of $\rho(770)$, we have only the virtual contribution for $K^+\bar{K}^0$ from the resonance in the decay 
$D^0 \to K^-\rho(770)^+ \to K^- K^+\bar{K}^0$. With the relation $g_{\rho(770)K^+ K^-}=g_{\omega(782)K^+ K^-}=
\frac{1}{\sqrt2}g_{\phi(1020)K^+ K^-}$~\cite{epjc39-41} for the strong couplings within flavour $SU(3)$ symmetry, the coefficient 
$c^K_{\rho(770)}=1.247\pm0.019$ was determined in Ref.~\cite{prd103-056021}, which is consistent with the results 
in~\cite{epjc39-41,prd81-094014,jetp129-386} for the kaon electromagnetic form factors. Then it's trivial to obtain the branching 
fraction 
\begin{eqnarray} 
  & & \mathcal{B}(D^0 \to K^-\rho(770)^+ \to K^- K^+\bar{K}^0)  \nonumber\\
  & & \hspace{1cm}    = (1.64\pm0.05\pm0.04\pm0.05)\times 10^{-4},           
   \label{res-D0K-R2KK}        
\end{eqnarray}
where the first two errors have the same sources as those in Eq.~(\ref{res-K-R+}), the third one comes from the uncertainty of 
coefficient $c^K_{\rho(770)}$. Considering the meson $\bar{K}^0$ decays half into ${K}^0_S$, one receives the contribution 
of the subprocess $\rho(770)^+\to K^+{K}^0_S$ to be about $(1.86\pm0.13\pm0.09)\%$ for the tree-body decay 
$D^0 \to K_S^0K^+K^-$ when comparing the result (\ref{res-D0K-R2KK}) with the corresponding data in Table~\ref{tab2}. 
The two errors come from the uncertainties of the data $(4.42\pm0.32)\times10^{-3}$ for $D^0 \to K_S^0K^+K^-$ and the 
result in Eq.~(\ref{res-D0K-R2KK}), respectively.

\begin{table}[thb] 
\begin{center}                                                                   
\caption{Experimental data of the branching fractions from {\it Review of Particle Physics}~\cite{PDG-2020} for the concerned 
              three-body $D$ meson decays.}
\label{tab2}   
\begin{tabular}{l c c } \hline\hline
     \quad Mode                                   &    \;\;Unit\;\;          & \;\;${\mathcal B}$~\cite{PDG-2020}  \\         \hline             
     $D^0 \to K_S^0K^+K^-$\;            &$\; 10^{-3} \;$       &\;$4.42\pm0.32$               \\ 
     $D^0 \to K_S^0K^-\pi^+$\;          &$\; 10^{-3} \;$       &\;$3.3\pm0.5$\;                   \\ 
     $D^0 \to K_S^0K^+\pi^-$\;          &$\; 10^{-3} \;$       &\;$2.17\pm0.34$                \\      
     $D^0 \to K^+K^-\pi^0$\;              &$\; 10^{-3} \;$       &\;$3.42\pm0.14$\vspace{0.1cm}  \\      
     $D^+ \to K^+K_S^0K_S^0$\;       &$\; 10^{-3} \;$       &\;$2.54\pm0.13$               \\ 
     $D^+ \to K^+K^+K^-$\;                &$\; 10^{-5} \;$       &\;$6.14\pm0.11$               \\      
     $D^+ \to K_S^0K^+\pi^0$\;         &$\; 10^{-3} \;$        &\;$5.07\pm0.30$               \\ 
     $D^+ \to K^+K^-\pi^+$\;              &$\; 10^{-3} \;$       &\;$9.68\pm0.18$\vspace{0.1cm} \\ 
     $D_s^+ \to K^+K^+K^-$\;            &$\; 10^{-4} \;$       &\;$2.16\pm0.20$                \\
     $D_s^+ \to K^+K^-\pi^+$\;           &$\; \% \;$              &\;$5.39\pm0.15$               \\      
     $D_s^+ \to K^+K_S^0\pi^0$\;      &$\; \% \;$              &\;$1.52\pm0.22$               \\       
\hline\hline
\end{tabular}   
\end{center}
\end{table}

\begin{figure}[tbp]  
\centerline{\epsfxsize=8.5cm \epsffile{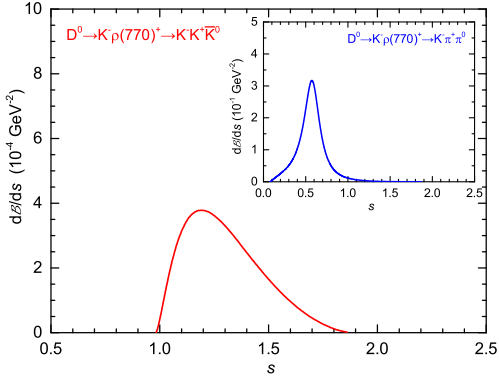}}
\vspace{-0.5cm}
\caption{The differential branching fractions for the quasi-two-body decays $D^0 \to K^-\rho(770)^+ \to K^- \pi^+\pi^0$ (inset)
               and $D^0 \to K^-\rho(770)^+ \to K^- K^+\bar{K}^0$.}
\label{fig-2}
\end{figure}

\begin{table*}[!]
\centering                                                                  
\caption{Virtual contributions of the subprocess $\rho(770)\to K\bar{K}$ for the concerned three-body $D$ decays.}
\label{tab-resv}   
\begin{tabular}{l c | c l } 
\hline\hline
            \;\;\; 2B  modes                                                        &  Data~\cite{PDG-2020}\quad\;                       
        & Q2B  modes \qquad\;\;                                                 &   \quad Results \;                                                     \\        
\hline             
          $D^0 \to K_S^0\rho(770)^0$\;                                   &$\;6.3^{+0.6}_{-0.8}\times10^{-3}\;$      
        &$D^0 \to K_S^0[\rho(770)^0\to]K^+K^-$                  &$\;2.48^{+0.24+0.10}_{-0.32-0.10}\times10^{-6}$    \\     
             ~                                                                              & ~      
        &$D^0 \to K_S^0[\rho(770)^0\to]K^0\bar K^0$          &$\;2.20^{+0.21+0.09}_{-0.28-0.09}\times10^{-6}$    \\               
          $D^0 \to K^-\rho(770)^+$\;                                       &$\;(11.3\pm0.7)\%\;$     
        &$D^0 \to K^-[\rho(770)^+\to]K^+\bar K^0$              &$\;1.71^{+0.11+0.07}_{-0.11-0.07}\times10^{-4}$   \\                
          $D^0 \to \pi^+\rho(770)^-$\;                                      &$\;(5.15\pm0.25)\times10^{-3}\;$     
        &$D^0 \to \pi^+[\rho(770)^-\to]K^- K^0$                    &$\;1.88^{+0.09+0.08}_{-0.09-0.08}\times10^{-5}$   \\      
          $D^0 \to \pi^-\rho(770)^+$\;                                      &$\;(1.01\pm0.04)\%\;$     
        &$D^0 \to \pi^-[\rho(770)^+\to]K^+\bar K^0$             &$\;3.69^{+0.15+0.15}_{-0.15-0.15}\times10^{-5}$   \\        
           $D^0 \to \pi^0\rho(770)^0$\;                                     &$\;(3.86\pm0.23)\times10^{-3}\;$     
        &$D^0 \to \pi^0[\rho(770)^0\to]K^+ K^-$                     &$\;3.69^{+0.22+0.15}_{-0.22-0.15}\times10^{-6}$   \\       
             ~                                                                               & ~      
        &$D^0 \to \pi^0[\rho(770)^0\to]K^0\bar K^0$             &$\;3.40^{+0.20+0.14}_{-0.20-0.14}\times10^{-6}$\vspace{0.15cm}  \\            
          $D^+ \to K_S^0\rho(770)^+$\;                                    &$\;6.14^{+0.60}_{-0.35}\%\;$      
        &$D^+ \to K_S^0[\rho(770)^+\to]K^+\bar K^0$           &$\;9.35^{+0.91+0.39}_{-0.53-0.39}\times10^{-5}$    \\               
          $D^+ \to K^+\rho(770)^0$\;                                        &$\;(1.9\pm0.5)\times10^{-4}\;$      
        &$D^+ \to K^+[\rho(770)^0\to]K^+K^-$                       &$\;7.81^{+2.06+0.32}_{-2.06-0.32}\times10^{-8}$    \\     
             ~                                                                                & ~      
        &$D^+ \to K^+[\rho(770)^0\to]K^0\bar K^0$                &$\;6.95^{+1.83+0.29}_{-1.83-0.29}\times10^{-8}$    \\     
          $D^+ \to \pi^+\rho(770)^0$\;                                       &$\;(8.3\pm1.5)\times10^{-4}\;$      
        &$D^+ \to \pi^+[\rho(770)^0\to]K^+K^-$                       &$\;7.99^{+1.44+0.33}_{-1.44-0.33}\times10^{-7}$    \\     
             ~                                                                                 & ~      
        &$D^+ \to \pi^+[\rho(770)^0\to]K^0\bar K^0$               &$\;7.36^{+1.33+0.30}_{-1.33-0.30}\times10^{-7}$\vspace{0.15cm}  \\                    
         $D_s^+ \to K^+\rho(770)^0$\;                                        &$\;(2.5\pm0.4)\times10^{-3}\;$      
        &$D_s^+ \to K^+[\rho(770)^0\to]K^+K^-$                      &$\;1.57^{+0.25+0.06}_{-0.25-0.06}\times10^{-6}$    \\     
             ~                                                                                  & ~      
        &$D_s^+ \to K^+[\rho(770)^0\to]K^0\bar K^0$              &$\;1.42^{+0.23+0.06}_{-0.23-0.06}\times10^{-6}$    \\                                                       
\hline\hline
\end{tabular} 
\end{table*}

In Fig.~\ref{fig-2}, we show the differential branching fractions for the decays $D^0 \to K^-\rho(770)^+ \to K^- \pi^+\pi^0$ and 
$D^0 \to K^-\rho(770)^+ \to K^- K^+\bar{K}^0$. In the inset, the feature of $\rho(770)$ is clearly and fully presented through the 
curve with a peak at about $s=m^2_{\rho(770)}$ for the decay process $D^0 \to K^-\rho(770)^+ \to K^- \pi^+\pi^0$. The obvious 
comparison is a bump at about $s=1.2$ GeV$^2$ for $D^0 \to K^-\rho(770)^+ \to K^- K^+\bar{K}^0$, which should not but 
potentially could be claimed as a resonant state with quite large decay width. The peak location of the bump for the subprocess 
$\rho(770) \to K \bar{K}$ of $D$ meson decay like $D^0 \to K^- K^+\bar{K}^0$ is distinctly different from it for the same subprocess 
in the three-body $B$ meson decays been studied in Refs.~\cite{prd101-111901,prd103-056021}, the former is closer to the pole mass 
of the resonance $\rho(770)$. This phenomenon is attributed to kinematic characteristics in the corresponding decay processes 
rather than the properties of the resonant states involved. Because the related three-body phase space of $D$ meson decays is 
much smaller than that of $B$ decays, the phase space factor $| \overrightarrow{p_h}|$ in Eq.~(\ref{eqn-diff-bra}) will drop
more quickly in the $D$ processes. This will also result in the ratio of the contributions from subprocesses $\rho(770)\to K\bar{K}$ 
to $\rho(770) \to \pi \pi$ to be channel-dependent in the three-body $D$ decays and much smaller than that for the three-body 
$B$ decays, which will decline from one or two percent in $B$ decays~\cite{prd103-056021} to the level of thousandth or even 
smaller, as exhibited by the results in Table~\ref{tab-resv}. 

The resonance $\rho(770)$ as a virtual bound state~\cite{Dalitz62,plb25-294} in the decay $\rho(770)\to K\bar K$ will not 
completely present its properties in the concerned processes because of the phase space. Nevertheless the quantum number 
of the involved resonance could be fixed from its decay daughters, for example $K\bar K$, along with CKM matrix elements in 
the decay on theoretical side. The certain resonant source for the final states $K\bar K$ makes the cascade decay like  
$D^0 \to K^-\rho(770)^+ \to K^- K^+\bar{K}^0$ to be a quasi-two-body process, although the  invariant mass region for the 
kaon pair is excluded from the region around pole mass of the intermediate state $\rho(770)$.  What we want to stress here is 
that the nonresonant contribution in the three-body $D$ or $B$ meson decays should not include the specific known contribution 
from a certain determinate resonant state like $\rho(770)$ for $K\bar K$ in the experimental studies. 

Different from the contribution of the subprocess $\rho(770)\to\pi\pi$ for the decays like $D^0 \to K^-\rho(770)^+$, the virtual 
contributions of the subprocesses $\rho(770)\to K\bar K$ for the concerned $D$ meson decays are nearly unaffected by the full 
width of the resonance $\rho(770)$. Take the decay $D^0 \to K^-\rho(770)^+ \to K^- K^+\bar{K}^0$ as an example, we will have 
its branching fraction slightly changed from $1.638\times 10^{-4}$ to $1.640\times 10^{-4}$ when the $\Gamma_{\rho(770)}$ is 
altered from $0.149$ GeV to zero. The best explanation for this goes to that the $s$-dependent width for BW formula of the 
Eq.~(\ref{eq-BW}) fades into insignificance when the invariant mass square $s$, which start from the threshold of kaon pair, is 
large enough, then the BW expression for $\rho(770)$ is charged only by the coefficient $c^K_\rho$ of the kaon form factors and 
the gap between the squared mass of the involved resonant state and the invariant mass square $s$ for kaon pair. Thanks to the 
phase space factor $| \overrightarrow{q}|^3$ in Eq.~(\ref{eqn-diff-bra}), the portion of the contribution in the shadow of the large 
decay width $\Gamma_{\rho(770)}$ for the process $D^0 \to K^-\rho(770)^+ \to K^- K^+\bar{K}^0$ is the strongly suppressed.

The branching fraction of the decay $D^0 \to K^-\rho(770)^+ \to K^- K^+\bar{K}^0$ can also be achieved from the 
Eq.~(\ref{rela-2B-Q2B}) with the help of the data for $D^0 \to K^-\rho(770)^+$. With the two-body branching fraction 
$\mathcal{B}(D^0 \to K^-\rho(770)^+)=(11.3\pm0.7)\%$~\cite{PDG-2020}, one has $\mathcal{B}=
(1.71\pm0.11\pm0.07)\times10^{-4}$ as shown in Table~\ref{tab-resv} for the quasi-two-body decay 
$D^0 \to K^-\rho(770)^+ \to K^- K^+\bar{K}^0$. This result is also in agreement with the value in Eq.~(\ref{res-D0K-R2KK}).
The first error of these results for the quasi-two-body decays in Table~\ref{tab-resv} comes from the uncertainties of the 
corresponding data for the relevant two-body decays in the same table, the second one is induced by the uncertainties of the 
coefficient $c^K_{\rho(770)}=1.247\pm0.019$~\cite{prd103-056021}, and decay the constant $f_\rho=0.216\pm0.003$ GeV~\cite{jhep1608-098}. 

The quasi-two-body branching fractions in Table~\ref{tab-resv},  which are derived from the experimental data for 
corresponding two-body decays with the Eq.~(\ref{rela-2B-Q2B}), are generally very small comparing with the corresponding  
experimental data in Table~\ref{tab2}, but some of them have the potential to be observed in the Dalitz plot analyses for related 
three-body $D$ decays. In addition to $D^0 \to K^-\rho(770)^+ \to K^- K^+\bar{K}^0$ with $\bar{K}^0\to{K}^0_S$, the 
subprocess $\rho(770)^+\to K^+{K}^0_S$ will contribute about $(1.84\pm0.09^{+0.19}_{-0.13})\%$ of the total branching 
fraction for the $D^+ \to K^+K_S^0K_S^0$ decay, with two errors come from the uncertainties of the data 
$\mathcal{B}=(2.54\pm0.13)\times10^{-3}$ in Table~\ref{tab2} and the corresponding quasi-two-body result in Table~\ref{tab-resv}.
While the quasi-two-body decays $D^0 \to \pi^-[\rho(770)^+\to]K^+K^0_S$ and $D_s^+ \to K^+[\rho(770)^0\to]K^+K^-$ are 
predicted to provide $(0.85\pm0.13\pm0.05)\%$ and $(0.73\pm0.07\pm0.12)\%$, respectively, for the total branching fractions 
of their corresponding three-body decay processes shown in Table~\ref{tab2}. 

Because of the phase space and the strong coupling, the virtual contribution of the subprocess $\rho(770)^0 \to K^+ K^-$ for 
the three-body decay $D^0 \to {K}^0_S K^+ K^-$ shown in Table~\ref{tab-resv} is very small, which is less than $1/10^3$ of the 
total branching fraction $(4.42\pm0.32)\times10^{-3}$ in Table~\ref{tab2} for the corresponding three-body decay process. 
Actually, because of the suppression from the factor $1/2$ in Eqs.~(\ref{def-F-u})-(\ref{def-F-d}) the kaon form factors, 
the subprocesses $\rho(770)^0 \to K^+ K^-$ and $\rho(770)^0 \to K^0 \bar K^0$ will contribute the smaller branching fractions 
comparing with the  $\rho(770)^\pm\to K^\pm K^0_S$. This partly causes the decays $D^0 \to K_S^0\rho(770)^0$,
$D^0 \to \pi^0\rho(770)^0$, $D^+ \to K^+\rho(770)^0$, and $D^+ \to \pi^+\rho(770)^0$ with $\rho(770)^0$ decaying into 
$K^+ K^-$ or $K^0 \bar K^0$ listed in Table~\ref{tab-resv} hold the very small proportions of the total branching fractions 
for their corresponding three-body decay processes, and are unlikely to be observed in the Dalitz plot analyses in the near future. 

With the $\omega$ components in Eqs.~(\ref{def-F-u})-(\ref{def-F-d}), the branching fractions for quasi-two-body decays with 
the intermediate state $\omega(782)$ for $K\bar K$ could be easily obtained from Eq.~(\ref{rela-2B-Q2B}) with the help of the existing 
experimental data. We predict the branching fractions for $D^0 \to K_S^0[\omega(782)\to]K^+K^-$ and $D_s^+ \to K^+[\omega(782)\to]K^+K^-$ to be $(4.59\pm0.25\pm0.40)\times10^{-6}$ and $(5.72\pm1.64\pm0.50)\times10^{-7}$, respectively, 
with the data $\mathcal{B}(D^0 \to K_S^0\omega(782))=(1.11\pm0.06)\%$ and $\mathcal{B}(D_s^+ \to K^+\omega(782)) =(8.7\pm2.5)\times10^{-4}$~\cite{PDG-2020}. And the first error comes from the uncertainty the corresponding data and second one 
is induced by $c^K_{\omega(782)}=1.113\pm0.019$~\cite{prd103-056021} and $f_\omega=0.197\pm0.008$ GeV~\cite{jhep1608-098}.

For the quasi-two-body decay $D^0 \to K^-\rho(1450)^+ \to K^- \pi^+\pi^0$, color-favored tree amplitude $T_P$ with 
$D^0\to K^-$ transition and W-exchange nonfactorizable amplitude $E_V$ are involved. Since the parameters $\chi^E_q=0.25$ 
and $\phi^E_q=1.73$ were fitted for the meson $\rho(770)$ in~\cite{prd89-054006}, one should not trust them in the decays with 
the subprocess $\rho(1450)^+ \to \pi^+\pi^0$. Fortunately, one has $T_P$ to be the dominated amplitude for this quasi-two-body
processes. By omitting the annihilation-type amplitude $E_V$, we estimate the branching fraction 
\begin{eqnarray} 
  \mathcal{B}  =   (2.25\pm0.56) \times 10^{-3}   
  \label{br-K-R1450+}  
\end{eqnarray}
for the $D^0 \to K^-\rho(1450)^+ \to K^- \pi^+\pi^0$ decay with the GS model measured in~\cite{prd86-032013} for the pion form 
factor, where only the error comes from the uncertainty of $|c^\pi_{\rho(1450)}|=0.158\pm0.018$~\cite{prd86-032013} has been 
taken into account.  If we switch on the amplitude $E_V$ and still adopt the parameters $\chi^E_q=0.25$ and 
$\phi^E_q=1.73$, we have $\mathcal{B} \approx 2.20\times 10^{-3}$ for this quasi-two-body process, which is very close to the 
value in Eq~(\ref{br-K-R1450+}). In Ref.~\cite{epjc39-41}, a smaller coefficient $c^\pi_{\rho(1450)}=-0.119\pm0.011$ was fitted for 
the pion form factor $F_\pi(s)$, with which a small branching fraction $\mathcal{B} \approx 0.85\times 10^{-3}$ could be obtained 
for the same quasi-two-body decay. In~\cite{PDG-2020,prd63-092001}, the branching fraction $(8.2\pm1.8)\times 10^{-3}$ was 
claimed for the quasi-two-body decay $D^0 \to K^-\rho(1700)^+ \to K^- \pi^+\pi^0$. While with the $F_\pi(s)$ measured
in~\cite{prd86-032013}, we estimate its branching fraction to be about $0.74 \times 10^{-3}$. It seems the result for this decay 
process in~\cite{prd63-092001} was overestimated.

\begin{figure}[tbp]  
\centerline{\epsfxsize=8.5cm \epsffile{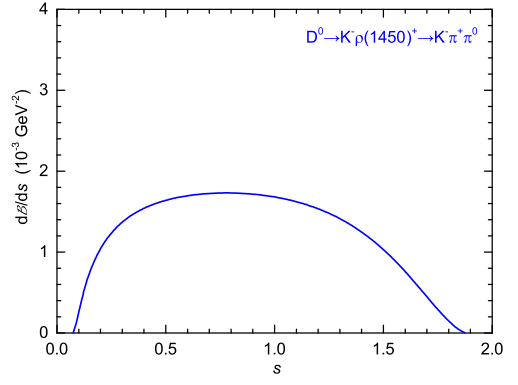}}
\vspace{-0.5cm}
\caption{The differential branching fractions for $D^0 \to K^-\rho(1450)^+ \to K^- \pi^+\pi^0$ decay.}
\label{fig-3}
\end{figure}

In Fig.~\ref{fig-3}, the differential branching fraction is shown for the decay $D^0 \to K^-\rho(1450)^+ \to K^- \pi^+\pi^0$, but 
one can not tell the feature of the resonance $\rho(1450)$ from the curve, it is more of a diagram for the background.
The reason for this is that the subprocess $\rho(1450)^+ \to \pi^+\pi^0$ in this decay is, in essence, a virtual resonance decay.
Although the pole mass of $\rho(1450)$ is larger than the threshold of pion pair, but the initial decaying state $D^0$ does not have 
the energy to make the resonance in $D^0 \to K^-\rho(1450)^+ \to K^- \pi^+\pi^0$ demonstrate its intact properties, it has already 
been terminated before the $s$ for the pion pair arrive the position of $m^2_{\rho(1450)}$. Contrary to the virtual contribution 
of $\rho(770) \to K \bar{K}$ discussed above, the contribution from $\rho(1450)^+ \to \pi^+\pi^0$ in the quasi-two-body decay 
$D^0 \to K^-\rho(1450)^+ \to K^- \pi^+\pi^0$ arises only from the forepart of BW expression for the involved resonance. 
Similar virtual contributions from the forepart of BW for the resonances will also take place in other decays and shall be leaved for 
future studies.
 
For the resonance $\rho(1450)$ decaying into kaon pair,  the $s$-dependent width $\Gamma_R(s)$ in Eq.~(\ref{eq-BW}) containing 
partial widths of various final states was adopted in Refs.~\cite{epjc79-436,epjc78-897,JHEP2001-112}.  In view of the decays 
$\rho(1450)\to \omega\pi$ and $\rho(1450)\to 4\pi$ are the two dominated modes for $\rho(1450)$, we adopt the 
$\Gamma_{\rho(1450)}(s)$ which is discussed in the Appendix with four decay channels.  
With Eq.~(\ref{G1450-4chs}), the coefficient $c^K_{\rho(1450)}=-0.156\pm0.015$~\cite{prd103-056021} and the amplitude $T_P$, 
it's easy to estimate the branching fraction
\begin{eqnarray} 
  &\mathcal{B}(D^0 \to K^-\rho(1450)^+ \to K^- K^+\bar K^0)  \nonumber\\
   & \qquad   =(0.91\pm0.17)\times 10^{-4},    
\end{eqnarray}
with the error comes from the uncertainty of $c^K_{\rho(1450)}=-0.156\pm0.015$~\cite{prd103-056021}. 
This result, just like the resonance contribution in the decay $D^0 \to K^-\rho(1450)^+ \to K^- \pi^+\pi^0$, 
is the virtual contribution from the forepart of BW formula of $\rho(1450)$.  And we need to stress here that, 
this virtual contribution from $\rho(1450)$ doesn't depend on the $\Gamma_{\rho(1450)}(s)$ in Eq.~(\ref{G1450-4chs}). 
Taking $\bar K^0\to K^0_S$ into account, the resonance  $\rho(1450)^+$ will contribute about $(1.03\pm0.19\pm0.07)\%$ for the 
total branching fraction of the tree-body decay $D^0 \to K_S^0K^+K^-$, where two errors come from the uncertainties the estimated 
branching fraction and the data $\mathcal{B}=(4.42\pm0.32)\times10^{-3}$ for the three-body decay, respectively.

When we put the decay amplitudes of two virtual contributions $\rho(770)^+\to K^+ K^0_S$ and 
$\rho(1450)^+\to K^+ K^0_S$ together for the tree-body decay $D^0 \to K_S^0K^+K^-$, we have the branching fraction
\begin{eqnarray} 
  &\mathcal{B}(D^0 \to K^-\rho(770,1450)^+ \to K^- K^+ K^0_S) \nonumber\\
   & \qquad    =(2.23\pm0.19)\times 10^{-4}.    
\end{eqnarray}          
If we turn on three resonances $\rho(770)^+,\rho(1450)^+$ and $\rho(1700)^+$,  the branching fraction will be enhanced
to be about $\mathcal{B}=3.54\times 10^{-4}$, which means about $8.0\%$, comparable to the result $22^{+8}_{-13}\%$ 
in~\cite{prd103-114028}, of the total branching fraction for $D^0 \to K_S^0K^+K^-$. Nevertheless, we need stress that there is no 
precise measurement for kaon form factors like $F_\pi(s)$ in~\cite{prd86-032013}, and the coefficients of $c^K_{\rho}$'s are not 
necessary to be real values, the phase difference between BW expressions of $\rho(770)^+,\rho(1450)^+$ and $\rho(1700)^+$ 
for $F_K(s)$ could change the weight of the interferences between them.

For the quasi-two-body decay $D^0 \to \pi^-\rho(1450)^+\to \pi^- K^+\bar K^0$, the decay of the subprocess 
$\rho(1450)^+\to K^+\bar K^0$ is an ordinary process because of $(m_{\pi^-}+m_{\rho(1450)})<m_{D^0}$. 
With Eq.~(\ref{G1450-4chs}),  one has its branching fraction
\begin{eqnarray} 
  &\mathcal{B}(D^0 \to \pi^-\rho(1450)^+\to \pi^- K^+\bar K^0) \nonumber\\
   & \qquad   =(3.72\pm0.71)\times 10^{-5},     
\end{eqnarray}
where the error comes from the uncertainty of $c^K_{\rho(1450)}=-0.156\pm0.015$.   
This result is close the virtual contribution from resonance $\rho(770)^+$
for the corresponding quasi-two-body decay process in the Table~\ref{tab-resv}. The other contributions of kaon pair from 
resonance $\rho(1450)$ for the concerned $D$ meson decays such as $D^0 \to \pi^-[\rho(1450)^+\to]K^+K^0_S$, the parameters 
like $D\to \rho(1450)$ transition form factors are absent in literature, we leave them for the future studies.

\section{Summary}
\label{sec-sum}

In this work, we studied the contributions of the subprocesses $\rho(770,1450)\to K\bar{K}$ for the three-body hadronic $D$ 
meson decays. We constructed the theoretical framework for the quasi-two-body $D$ decays with the help of the pion and 
kaon electromagnetic form factors and derived the relation connecting the branching fractions for the two-body with that for the  
corresponding quasi-two-body $D$ meson decays. With which we obtained the numerical results for the concerned 
quasi-two-body $D$ meson decay processes within the flavour $SU(3)$ symmetry.

We predicted the branching fraction 
$\mathcal{B}=(0.82\pm0.04)\times 10^{-4}$ for the decay $D^0 \to K^-\rho(770)^+ \to K^- K^+K_S^0$, which is $(1.86\pm0.16)\%$
of the total branching ratio $(4.42\pm0.32)\times10^{-3}$ in \emph{Review of Particle Physics} for the three-body decay 
$D^0 \to K_S^0K^+K^-$. While the subprocess $\rho(1450)^+\to K^+\bar K^0$ was found to contribute a branching fraction 
$\mathcal{B}=(0.91\pm0.17)\times 10^{-4}$ for the quasi-two-body decay $D^0 \to K^-\rho(1450)^+ \to K^- K^+\bar K^0$ in this work. The subprocess $\rho(770)^+\to K^+{K}^0_S$, in addition, could contribute about $(1.84^{+0.21}_{-0.16})\%$ of the total branching 
fraction for the $D^+ \to K^+K_S^0K_S^0$ decay. And the quasi-two-body decays $D^0 \to \pi^-[\rho(770)^+\to]K^+K^0_S$ and 
$D_s^+ \to K^+[\rho(770)^0\to]K^+K^-$ were predicted to provide $(0.85\pm0.14)\%$ and $(0.73\pm0.14)\%$, respectively, for 
the total branching fractions of their corresponding three-body decay processes in this paper, which have the potential to be 
observed in the Dalitz plot analyses for related three-body $D$ decays.

Different from $\rho(770)\to\pi\pi$, the subprocess $\rho(770)\to K\bar K$ will provide only the virtual 
contribution for the concerned three-body $D$ meson decays in this work. And the virtual bound state $\rho(770)$ decaying 
into kaon pair can not completely present its properties in relevant decay processes because of the phase space limitation.  
The situation for $\rho(1450)^+$ which is decaying into $\pi^+\pi^0$ or $K^+\bar{K}^0$ 
in $D^0 \to K^-\rho(1450)^+$  is similar with that of $\rho(770)$ decaying to $K\bar K$.  Although the pole mass of 
$\rho(1450)$ is larger than the threshold of pion and kaon pairs, but the initial decaying state $D^0$ does not have enough energy 
to make this resonance demonstrate its intact properties in $D^0 \to K^-\rho(1450)^+$. The virtual contributions from various 
resonant states are widely exists in multibody $D$, $J/\psi$, $B$, etc. decays.  We need to stress that the virtual 
contributions from specific known intermediate states is different from the nonresonant contributions demarcated in the 
experimental studies. 

\begin{acknowledgments}
We thank Profs. Hsiang-nan Li and Cai-Dian L\"u for valuable discussions. 
This work was supported in part by the National Natural Science Foundation of China 
        under Grants No.~11547038  and No.~12047571.
\end{acknowledgments}

\appendix
\section{$\Gamma_R(s)$ for $\rho(1450)$} 
\label{sec-appx}

Different from the BW formula for $\rho(770,1450)$ in~\cite{prl123-231802,epjc39-41,prd81-094014,prd86-032013}, 
the $s$-dependent width $\Gamma_R(s)$ in Eq.~(\ref{eq-BW}) containing partial widths of various final states for resonances 
$\rho(770,1450)$ was adopted in Refs.~\cite{epjc79-436,epjc78-897,JHEP2001-112}. 
Since $\mathcal{B}(\rho(770)\to \pi\pi)\sim 100\%$~\cite{PDG-2020}, $\Gamma_{\rho(770)}(s)=\Gamma_{\rho(770)\to\pi\pi}(s)$ 
is a very good approximation for the BW formula for its $\pi\pi$ and $K\bar K$ final states, besides, the virtual contributions of 
the subprocesses $\rho(770)\to K\bar K$ are unaffected by the full width of $\rho(770)$ for the concerned $D$ meson 
decays as discussed in Sec.~\ref{sec-3}.

For the $\Gamma_R(s)$ in BW formula of resonance $\rho(1450)$, the $s$-dependent width with two channels
\begin{eqnarray}
  && \Gamma_{\rho(1450)}(s)=\Gamma_{\rho(1450)}\frac{m_R}{\sqrt s}  \nonumber\\
  &&\hspace{0.5cm}  \times\left[ \frac{ \left| \overrightarrow{q} \right|^3}{ \left| \overrightarrow{q_{0}}\right|^3} \theta(s-4m^2_\pi)
                            +  \frac12 \frac{ \left| \overrightarrow{q_K} \right|^3}{ \left| \overrightarrow{q_{K0}}\right|^3} \theta(s-4m^2_K)\right]\!, 
                            \quad
\end{eqnarray}
 and
\begin{eqnarray}
    \Gamma_{\rho(1450)}(s)=\Gamma_{\rho(1450)}\frac{m_R}{\sqrt s} 
     \frac{2 \left| \overrightarrow{q} \right|^3 +\left| \overrightarrow{q_K} \right|^3 }{ 2\left| \overrightarrow{q_{0}}\right|^3},
\end{eqnarray}
were adopted in~\cite{epjc79-436} and~\cite{epjc78-897}, respectively, where $\overrightarrow{q}_{K(0)}$ is the 
$\overrightarrow{q}_{(0)}$ with the replacement $m_\pi\to m_K$. In view of that the decays $\rho(1450)\to \omega\pi$ and 
$\rho(1450)\to 4\pi$ are the two dominated modes for the resonance $\rho(1450)$, we adopt the $\Gamma_{\rho(1450)}(s)$ 
with four channels in this work as~\cite{JHEP2001-112,prd77-092002}
\begin{eqnarray}
&&   \Gamma_{\rho(1450)}(s) =\Gamma_{\rho(1450)}  \frac{m_R}{\sqrt s}  \nonumber\\ 
&& \hspace{1cm} \times\bigg[  \mathcal{B}(\rho(1450)\to \pi\pi)  \frac{ \left| \overrightarrow{q} \right|^3}
                                { \left| \overrightarrow{q_{0}}\right|^3}    X^2(\left| \overrightarrow{q} \right| r^R_{\rm BW})  \nonumber\\ 
&& \hspace{1cm}+ \mathcal{B}(\rho(1450)\to K\bar K)  \frac{ \left| \overrightarrow{q_K} \right|^3}{ \left| \overrightarrow{q_{K0}}\right|^3}   
                             X^2(\left| \overrightarrow{q_K} \right| r^R_{\rm BW})  \nonumber\\                     
&& \hspace{1cm}+ \mathcal{B}(\rho(1450)\to \omega\pi) \frac{ \left| \overrightarrow{q_\omega} \right|^3}
                     { \left| \overrightarrow{q_{\omega0}}\right|^3}    X^2(\left| \overrightarrow{q_\omega} \right| r^R_{\rm BW})  \nonumber\\                                          
&& \hspace{1cm}+\mathcal{B}(\rho(1450)\to 4\pi) \frac{ \left| \overrightarrow{q_{4\pi}} \right|^3}
                  { \left| \overrightarrow{q_{4\pi0}}\right|^3}      X^2(\left| \overrightarrow{q_{4\pi}} \right| r^R_{\rm BW})  \bigg]\!,\quad
   \label{G1450-4chs}
\end{eqnarray}
where~\cite{JHEP2001-112} 
\begin{eqnarray}    
   &&\left| \overrightarrow{q_\omega} \right|=                        \nonumber \\
     && \quad\frac{1}{2\sqrt s} \sqrt{\left[s-(m_\omega+m_{\pi})^2\right]\left[s-(m_\omega-m_{\pi})^2\right]}, \;\;\; 
                \label{def-qh}
\end{eqnarray}  
and $\left| \overrightarrow{q_{4\pi}} \right|=\frac12\sqrt{s-16m^2_\pi}$~\cite{prd77-092002}, 
and  $\left| \overrightarrow{q_{\omega0}}\right|$ and $\left| \overrightarrow{q_{4\pi0}} \right|$ are the 
$\left| \overrightarrow{q_{\omega}}\right|$ and $\left| \overrightarrow{q_{4\pi}} \right|$ at $s=m^2_{\rho(1450)}$, respectively. 
We have the branching fractions $\mathcal{B}(\rho(1450)\to \omega\pi)=45\%$ and 
$\mathcal{B}(\rho(1450)\to 4\pi)=40\%$~\cite{JHEP2001-112}, and 
$\mathcal{B}(\rho(1450)^0\to K^+ K^-)/\mathcal{B}(\rho(1450)^0\to \pi^+ \pi^-)\approx1/10$~\cite{prd101-111901}.


\end{document}